\documentclass[utf8]{frontierarxiv} % for Physics and Applied Mathematics and Statistics articles
\usepackage{url,lineno,microtype,subcaption}
\usepackage[onehalfspacing]{setspace}

\def\keyFont{\fontsize{8}{11}\helveticabold }
\def\firstAuthorLast{Hömmen and Mäkinen {et~al.}} %use et al only if is more than 1 author
\def\Authors{Peter Hömmen\,$^{+,1,*}$, Antti J. Mäkinen\,$^{+,2,*}$, Alexander Hunold\,$^{3}$, René Machts\,$^{3}$, Jens Haueisen\,$^{3}$, Koos C. J. Zevenhoven\,$^{2}$, Risto J. Ilmoniemi\,$^{2}$, and Rainer Körber\,$^{1}$}

\def\Address{$^{+}$These authors contributed equally to this work\\
$^{1}$Physikalisch-Technische Bundesanstalt, Berlin, Germany\\
$^{2}$Aalto University School of Science, Department of Neuroscience and Biomedical Engineering, Espoo, Finland\\
$^{3}$Technische Universität Ilmenau, Institute of Biomedical Engineering and Informatics, Ilmenau, Germany}

\def\corrAuthor{Peter Hömmen\\ Physikalisch-Technische Bundesanstalt (PTB), Abbestrasse 2-12, 10587 Berlin, Germany}

\def\corrEmail{peter.hoemmen@ptb.de}

\usepackage[utf8]{inputenc}
\usepackage{amsmath}
\usepackage{amsfonts}
\usepackage{amssymb}
\usepackage{graphicx}
\usepackage{float}
\usepackage{authblk}

\usepackage{upgreek}

\begin{document}
\twocolumn
\firstpage{1}

\title[Performance of ULF MRI for CDI of the Human Head]{Evaluating the Performance of Ultra-Low-Field MRI for \textit{In-vivo} 3D Current Density Imaging of the Human Head} 

\author[\firstAuthorLast ]{\Authors} %This field will be automatically populated
\address{} %This field will be automatically populated
\correspondance{} %This field will be automatically populated

%\extraAuth{}% If there are more than 1 corresponding author, comment this line and uncomment the next one.
\extraAuth{Antti J. Mäkinen \\ Aalto University School of Science, Department of Neuroscience and Biomedical Engineering, Rakentajanaukio 2B, 02150 Espoo, Finland, antti.makinen@aalto.fi}

\maketitle
%\maketitle
\begin{abstract}
\section{}
\noindent
Magnetic fields associated with currents flowing in tissue can be measured non-invasively by means of zero-field-encoded ultra-low-field magnetic resonance imaging (ULF MRI) enabling current density imaging (CDI) and possibly conductivity mapping of human head tissues. Since currents applied to a human are limited by safety regulations and only a small fraction of the current passes through the relatively high-resistive skull, a sufficient signal-to-noise ratio (SNR) may be difficult to obtain when using this method. In this work, we study the relationship between the image SNR and the SNR of the field reconstructions from zero-field-encoded data. We evaluate these results for two existing ULF MRI scanners---one ultra-sensitive single-channel system and one whole-head multi-channel system---by simulating sequences necessary for current-density reconstruction. We also derive realistic current-density and magnetic-field estimates from finite-element-method simulations based on a three-compartment head model. We found that existing ULF-MRI systems reach sufficient SNR to detect intra-cranial current distributions with statistical uncertainty below 10\%. However, they also reveal that image artifacts influence the reconstruction quality. Further, our simulations indicate that current-density reconstruction in the scalp requires a resolution less than 5 mm and demonstrate that the necessary sensitivity coverage can be accomplished by multi-channel devices. 
\tiny
 \keyFont{ \section{Keywords:} ultra-low-field MRI, current-density imaging, zero-field encoding, signal-to-noise ratio, finite-element method, Monte-Carlo simulation, MRI simulation} %All article types: you may provide up to 8 keywords; at least 5 are mandatory.
\end{abstract}

\section{Introduction}
Imaging of current-density distributions, produced by injecting current \textit{in vivo} into the human head, has a variety of possible applications. Three-dimensional conductivity distributions or simplified conductivity models may be extracted from such images. These are required for accurate source estimation in electromagnetic neuroimaging \cite{Haemaelaeinen1993, vallaghe2008}. Further, individual conductivity information is necessary for models used to optimize and plan therapeutic treatments, \textit{e.g.}, in transcranial magnetic (TMS) \cite{opitz2011how, nummenmaa2013comparison} and transcranial direct-current stimulation (tDCS) \cite{Miranda2018}. In addition, the current flow during tDCS may be monitored online, providing direct feedback.

Magnetic resonance imaging (MRI) is affected by local magnetic fields, such as the magnetic field $\boldsymbol{B}_J(\boldsymbol{r})$ associated with a current density $\boldsymbol{J}(\boldsymbol{r})$ at points $\boldsymbol{r}$ in the imaging volume. In particular, if also the main magnetic field $B_0$ can be switched on and off during the pulse sequence \cite{zevenhoven2014ultra, Hoemmen2019}, it is possible to measure full-tensor information of the effects of $\boldsymbol{B}_J(\boldsymbol{r})$, providing a way to directly estimate $\boldsymbol{J}(\boldsymbol{r})$ \cite{Vesanen2014, Nieminen2014}. Zero-field-encoded current density imaging (CDI), proposed by Vesanen et al. \cite{Vesanen2014}, has recently been demonstrated in phantom measurements and is most promising regarding {\it in-vivo} implementation \cite{Hoemmen2019}. This is made possible by superconducting quantum interference device (SQUID)-based ultra-low-field (ULF) MRI. Since current impressed \textit{in vivo} in the human head is limited by safety regulations to the low-mA range \cite{Bikson2016, Antal2017} and only small fraction of the current passes the relatively high-resistive skull \cite{Miranda2006, Neuling2012}, a sufficient signal-to-noise ratio (SNR) may be difficult to reach. 

%Magnetic resonance imaging (MRI) can be used for non-invasive current density imaging (CDI). Here, the changes in magnitude and phase of the MR signal due to the magnetic field $\boldsymbol{B}_J$, produced by the underlying current density $\boldsymbol{J}$, are detected. In contrast to methods that employ conventional (high-)field MRI machines, superconducting quantum interference device (SQUID)-based ultra-low-field (ULF) MRI offers the possibility to derive the full tensor of $\boldsymbol{B}_J$ \cite{Vesanen2014, Nieminen2014}, providing a direct method to estimate $\boldsymbol{J}$. Zero-field encoding as proposed by Vesanen \textit{et al.} \cite{Vesanen2014} has recently been demonstrated in phantom measurements and is most promising regarding \textit{in-vivo} implementation \cite{Hoemmen2019}. Since current impressed \textit{in vivo} in the human head is limited by safety regulations to the low mA range \cite{Bikson2016, Antal2017} and only small fraction of the current passes the relatively high-resistive skull \cite{Miranda2006, Neuling2012}, a sufficient signal-to-noise ratio (SNR) may be difficult to reach. 

The two main factors influencing the SNR in ULF MRI are system noise and the strength of the polarizing field that creates the necessary sample magnetization. Both issues have been addressed in previous setups. However, the ultimate sensitivity combining the lowest noise and the highest polarizing field in a single setup has not been demonstrated. Hömmen \textit{et al.} used an ultra-sensitive single-channel SQUID system with a noise level of 380~$\text{aT/}\sqrt{\text{Hz}}$ for the demonstration of CDI \cite{Hoemmen2019}. This noise performance was about 10--20 times better than in commercially available SQUID systems, but the polarizing field of 17~mT was comparatively low. Other groups reported ULF--MRI systems with polarizing fields over 100~mT, using cooled copper-coil setups \cite{Espy2015, Inglis2013}. Even higher polarizing fields could be reached by means of superconducting polarizing coils as presented in \cite{Vesanen2013, Lehto2017}.  

A quantitative survey of the necessary SNR for zero-field-encoded CDI with a defined uncertainty is still pending. In this work, we investigate the influence of noise on the quality of the $\boldsymbol{B}_{J}$ and $\boldsymbol{J}$ reconstructions by analytic approximations and by means of Monte-Carlo simulations. Our results enable the estimation of the required image SNR for a given statistical uncertainty in the field reconstructions. They further provide an intuitive method to assess the performance of a specific system for current-density imaging.

In addition, two existing ULF--MRI setups are examined more closely regarding their performance in a CDI application. The first is the single-channel setup of PTB, Berlin, described in \cite{Hoemmen2019}, which is now equipped with an updated polarization setup specially designed for the shape of the human head. The second setup is a whole-head multi-channel system, a successor of \cite{Vesanen2013}, located at Aalto University, Helsinki. The latest version comprises an optimized superconductive polarizing coil \cite{Lehto2017}, an ultra-low-noise amplifier for for flexible switching of all MRI fields \cite{zevenhoven2014ultra}, and newly developed SQUID-sensors specially designed for pulsed-field applications \cite{luomahaara2018}. 

Realistic $\boldsymbol{B}_{J}$ and $\boldsymbol{J}$ distributions were derived from  finite-element-method (FEM) simulations using a three-compartment head model. Combined with nominal gradient fields and sensitivity parameters of the described setups, the $\boldsymbol{B}_{J}$ distributions were put into a Bloch equation solver that emulates complete gradient-echo sequences in the time domain. Our simulation results not only provide a good estimate of the statistical uncertainty in zero-field-encoded CDI with currently available technologies but also reveal other important requirements in terms of sample coverage and image resolution.

\section{Zero-field-encoded CDI}
\label{Sec:State of the art}
To understand the effects of noise, we recap the sequence and reconstruction method designed by Vesanen \textit{et al.}~\cite{Vesanen2014}. At first, magnetization is built up by a polarization period. Subsequently, all MRI fields are turned off and the current density $\boldsymbol{J}$ is applied during a defined zero-field time $\tau$. After the zero-field time, the magnetization has been rotated to $\boldsymbol{m}_{1}$ by the magnetic field during $\tau$ as
\begin{equation}
\label{eq:Rotation}
    \boldsymbol{m}_{1}(\boldsymbol{r})=e^{\gamma\tau\mathbf{A}(\boldsymbol{r})}\boldsymbol{m}_{0}(\boldsymbol{r})=\mathbf{\Phi}(\boldsymbol{r})\boldsymbol{m}_{0}(\boldsymbol{r})\,,
\end{equation}
where, $\boldsymbol{m}_{0}$ is the starting magnetization and $\gamma$ the gyromagnetic ratio of the proton. $\mathbf{A}$ is a generator to the rotation matrix $\boldsymbol{\Phi}$, which describes the spin dynamics due to the quasi-static magnetic field during $\tau$ \cite[p.~86--89]{Kraus2014, Vesanen2014}. Ideally, this field is solely determined by the magnetic field $\boldsymbol{B}_{J}$ associated with $\boldsymbol{J}$. In reality, a superposition of a static background field and transient fields due to pulsing (in the following combined in the term $\boldsymbol{B}_\text{B}$) are present. Hence, the time evolution of $\boldsymbol{m}$ is affected by $\mathbf{A}_{J}+\mathbf{A}_{\text{B}}$ where $\mathbf{A}_{J}$ and $\mathbf{A}_{\text{B}}$ are associated with $\boldsymbol{B}_{J}$ and the average $\boldsymbol{B}_{\text{B}}$, respectively.

Following $\tau$, the main field $\boldsymbol{B}_{\text{0}}$, here in the \textit{x}-direction, is turned on and the magnetization is manipulated by gradient fields to encode spatial information in the phase and frequency of the resulting signal. Ignoring relaxation, the magnetic signal recorded at a sensor during the echo can be written as
\begin{equation}
\label{eq:Spatial encoding}
\begin{split}
    S(t) 
    &=\int \boldsymbol{C}(\boldsymbol{r})^{\top} \boldsymbol{m}(\boldsymbol{r}, t)dV  \\
    &= \int \boldsymbol{C}(\boldsymbol{r})^{\top} \mathbf{R}_{\text{f}}(\boldsymbol{r}, t) \mathbf{R}_{\text{p}}(\boldsymbol{r}) \boldsymbol{m}_{\text{1}}(\boldsymbol{r})dV\,,    
\end{split}
\end{equation}
where $t$ is the time, $\boldsymbol{C}$ the coupling field of the sensor, and matrices $\mathbf{R}_{\text{f}}$ and $\mathbf{R}_{\text{p}}$ correspond to rotations in the \textit{yz} plane related to the frequency and phase encoding parameters. For the following operations, it is convenient to convert the signal equation to a complex representation. Considering only the frequency components close to the Larmor frequency $\gamma |\boldsymbol{B}_{\text{0}}|$, the signal can be written as \cite{zevenhoven2019superconducting, brown2014}
\begin{equation}
\label{eq:complex_signal}
    S(t) \approx \operatorname{Re} \int \beta(\boldsymbol{r})^* e^{i[\omega(\boldsymbol{r}) t + \theta_\mathrm{p}(\boldsymbol{r})]}\Tilde{m}_{1}(\boldsymbol{r})dV\,,
\end{equation}
where $\beta = C_{z} + i C_{y}$, $\Tilde{m}_{1} = m_{\text{1},z} + i m_{\text{1},y}$, $\omega t$ is the phase angle due to precession during frequency encoding, and $\theta_\mathrm{p}$ the angle due to phase encoding. In a realistic setting, $\beta$ could also include additional effects from an inhomogeneous polarizing field and non-idealities in field pulsing.

After applying the discrete Fourier transform to the frequency- and phase-encoded data and taking the relevant frequency bins, the magnitude and phase of the rotation of $\boldsymbol{m}$ can be estimated at the location of the corresponding voxel. The voxel value corresponding to the MR signal generated close to $\boldsymbol{r}_n$ is given by
\begin{equation}
\label{eq:srf}
\begin{split}
    v_{n} 
    &= \int \text{SRF}(\boldsymbol{r}-\boldsymbol{r}_n) \beta(\boldsymbol{r})^* \Tilde{m}_{1}(\boldsymbol{r})dV \\
    &\approx \beta^* (\boldsymbol{r}_{n})\Tilde{m}_{1}(\boldsymbol{r}_{n})\,,
\end{split}
\end{equation}
where $\text{SRF}_{n}(\boldsymbol{r})$ is the spatial response function of the $n^\text{th}$ voxel \cite{Pruessmann2006}. When the SRF is close to a delta function $\delta(\boldsymbol{r} - \boldsymbol{r}_{n})$, the integral can be approximated with the function value at $\boldsymbol{r}_{n}$, otherwise the SRF will result in leakage artifacts from the neighbouring areas.

The voxel values $v_{n}$ contain information about the zero-field-encoded magnetic field in both their magnitude and phase. In reality, there are other factors, such as non-idealities in the gradient ramps and unknown relaxation profiles, that affect the voxel values as well. Therefore, the relative changes in $v_{n}$ associated with the current density are recovered by normalization with a reference $u_{n}$ \cite{Vesanen2014}, \cite{Hoemmen2019}. Repeating the sequence for all the three basis directions $\boldsymbol{e}_x$, $\boldsymbol{e}_y$, and $\boldsymbol{e}_z$, the last two rows of $\boldsymbol{\Phi}_{n}$ can be measured. For example, the \textit{y} and \textit{z} elements of the first column are given by:
\begin{equation}
\label{eq:Phi extraction}
\begin{split}
\Phi_{n(31)}&=\operatorname{Re}[v_{n,x}/u_{n}]\\     
\Phi_{n(21)}&=\operatorname{Im}[v_{n,x}/u_{n}]\,,
\end{split}
\end{equation}
where $v_{n,x}$ denotes the voxel value of a zero-field-encoded image with starting magnetization in the $x$ direction.
Rotation matrices are orthogonal by definition. Therefore, the first row of $\mathbf{\Phi}_{n}$ can be derived by the cross product of the second with the third row. Naturally, the elements in $\mathbf{\Phi}_{n}$ are contaminated by noise. A practical approach to increase the accuracy is to apply an orthogonalization. For this purpose Vesanen \textit{et al.}~\cite{Vesanen2014} suggest Löwdin's transformation, which yields the closest orthogonalization in the least-squares sense \cite{Loewdin1950}, \cite{Aiken1980}. It is clear that a unique rotation matrix $\boldsymbol{\Phi}_{n}$ is created for each voxel $n$. The following analysis in this section and in Sec.~\ref{Noise in zCDI} concentrates on a voxel-wise reconstruction of $\boldsymbol{B}_{J}$ and $\boldsymbol{J}$, where the index $n$ is left out for simplicity. 

Using $\boldsymbol{\Phi}$, all components of the magnetic field $\boldsymbol{B}=\boldsymbol{B}_{\text{B}}+\boldsymbol{B}_{J}$ can be derived from a non-linear inversion of the matrix exponential: 
\begin{equation}
\label{eq:Inversion}
\begin{split}
\gamma\tau\mathbf{A} 
&=
\gamma\tau\begin{bmatrix}
 0 & \hat{B}_{z} & -\hat{B}_{y}\\
 -\hat{B}_{z} & 0 & \hat{B}_{x}\\
 \hat{B}_{y} & -\hat{B}_{x} & 0
 \end{bmatrix}\\\\
 &= \frac{\phi}{2\sin\phi} (\boldsymbol{\Phi} - \boldsymbol{\Phi}^{\top})\,,
\end{split}
\end{equation} 
where $\phi = \arccos[(\operatorname{tr}(\boldsymbol{\Phi})-1)/2]$ represents the rotation angle of $\boldsymbol{\Phi}$ \cite{Vesanen2014}, and $\boldsymbol{\hat{B}}$ is the reconstruction of $\boldsymbol{B}$. From here on, reconstructed quantities are denoted using the hat symbol.

Finally, $\boldsymbol{\hat{B}}_J$ and $\boldsymbol{\hat{B}}_\text{B}$ can be decomposed from $\boldsymbol{\hat{B}}$ by the subtraction of another reconstruction. This could be a full 3D image of $\boldsymbol{B}_\text{B}$ only, or of $\boldsymbol{B}_\text{B}+\boldsymbol{B}_J$, with the impressed current having the opposite polarity. The latter reduces the statistical uncertainty by $1/\sqrt{2}$ and is from here on called bipolar reconstruction:
\begin{equation}
\label{eq:bipolar}
\begin{split}
\boldsymbol{\hat{B}}_{J} &= \dfrac{\boldsymbol{\hat{B}}_1-\boldsymbol{\hat{B}}_2}{2}\,,\\
\\
\boldsymbol{\hat{B}}_{1} &= \boldsymbol{\hat{B}}_{\text{B}}+\boldsymbol{\hat{B}}_{J(+)}\,,\\
\boldsymbol{\hat{B}}_{2} &= \boldsymbol{\hat{B}}_{\text{B}}+\boldsymbol{\hat{B}}_{J(-)}\,.
\end{split}
\end{equation} 
From Eq.~\eqref{eq:bipolar}, the full tensor of the local field $\boldsymbol{\hat{B}}_J$ is derived, enabling the estimation of $\boldsymbol{\hat{J}}$ by Ampère's Law:

\begin{equation}
\label{eq:Amperes_Law}
\boldsymbol{\hat{J}}=\frac{1}{\mu_{0}} \nabla \times \boldsymbol{\hat{B}}_{J}\,,
\end{equation}  
where $\mu_{0}$ is the permeability of free space.

\section{Noise in Zero-Field CDI}
\label{Noise in zCDI}
\subsection{The connection between noise in $\Phi$ and image SNR}
In this section, we analyze how the uncertainty in the reconstruction of zero-field-encoded data relates to the image SNR. From Eq.~\eqref{eq:Phi extraction}, we know that the values in $\mathbf{\Phi}$ are normalized by a complex reference $u=|u|e^{i\delta}$, where $|u|$ is related to the magnitude of the magnetization after $\tau$ and $\delta$ to the phase accumulation due to effects that do not arise from $\boldsymbol{B}_{J}+\boldsymbol{B}_{\text{B}}$.

Hömmen \textit{et al.}~\cite{Hoemmen2019} describe that $|u|$ cannot be measured directly due to the always present background field. However, the reference can be constructed from the real or imaginary parts of the three measurements of $v$ by
\begin{equation}
    \label{refernce}
    |u| = \sqrt{\operatorname{Re}[v_x]^{2}+\operatorname{Re}[v_y]^{2}+\operatorname{Re}[v_z]^{2}}\,,
\end{equation}
which effectively normalizes the rows of $\boldsymbol{\Phi}$ to exactly unit norm. The reference phase $\delta$, on the other hand, has to be acquired in a separate measurement. See \cite{Hoemmen2019} for more detail.

The complex reference value can be modeled as $u = \operatorname{E}[u] + \epsilon$, where $\operatorname{E}$ denotes the expected value and $\epsilon \sim \mathcal{N}(0, \sigma^2)$ is symmetric complex Gaussian noise that can be extracted from a noise-only image $e$, or from a noise-only region in any of the images $v$. Using this reference, we define the image SNR as
\begin{equation}
\begin{split}
    \mathrm{SNR} &\overset{\mathrm{def}}{=} \frac{|\operatorname{E}[u]|}{\mathrm{SD}[e]}
    = \frac{|\operatorname{E}[u]|}{\sqrt{\operatorname{E}[\operatorname{Re}(\epsilon)^2] + \operatorname{E}[\operatorname{Im}(\epsilon)^2]}} \\
    &= \frac{|\operatorname{E}[u]|}{\sigma}\,,
\end{split}
        \label{eq:SNRdef}
\end{equation}
where $\operatorname{SD}$ is the standard deviation. 

The phase correction with the noisy reference phase $\delta$ causes the real part to leak to the imaginary part and vice versa, increasing the noise in the matrix elements. Dividing by the magnitude of the complex reference $u = |u|e^{i\delta}$ yields unit norm in the rows of $\boldsymbol{\Phi}$ decreasing the noise. This is derived in the Appendix, which also shows that the noise SD in the elements of $\boldsymbol{\Phi}$ can be approximated as 
%\begin{equation}
%\label{eq:SNR}
%\begin{split}
%\sigma_{\Phi_{31}} 
%& = \operatorname{SD}[ {\operatorname{Re}(v(\boldsymbol{e}_{x})/u)}]\\ 
%& =\frac{1}{\sqrt{2}\,\mathrm{SNR}}g_{31}(\Phi)\\
%& \approx \frac{1}{\sqrt{2}\,\mathrm{SNR}}\sqrt{\frac{|\operatorname{E}[v(\boldsymbol{e}_{x})]\%,|^2}{|\operatorname{E}[u]\,|^2}+1},
%\end{split}
%\end{equation}
\begin{equation}
\sigma_{\Phi_{ij}} =\frac{1}{\sqrt{2}\,\mathrm{SNR}}g_{ij}(\boldsymbol{\boldsymbol{\Phi}})\,,
\label{eq:SNR}
\end{equation} 
where the scaling $1 \leq g_{ij}(\boldsymbol{\Phi}) \leq \sqrt{2}$ depends on the associated measurement. This approximation is valid when $u \approx \operatorname{E}[u]$, \textit{i.e.}, $\operatorname{SNR} \gg 1$. Eq.~\eqref{eq:SNR} already gives an impression of the noise SD in $\boldsymbol{\Phi}$ as a function of the image SNR. The rotation-dependent scaling $g_{ij}(\boldsymbol{\Phi})$ and correlations between the elements are given in the Appendix. 

The most important factors determining the SNR are the polarizing field, the coupling to the sensors, and the relaxation of the magnetization, all of which affect the voxel magnitude. The noise in the voxel values is governed by the system noise determined by the magnetic sensor as well as other instrumentational and environmental noise sources.

\subsection{Noise analysis of B-field reconstruction: linear approximation}
\label{sec:linear_approx}
To estimate the noise in the reconstruction of $\boldsymbol{B}$, we first discuss an idealized case, where all three rows of $\boldsymbol{\Phi}$ can be measured and no reference image $u$ is needed. In this case, the noise in the elements of $\boldsymbol{\Phi}$ becomes independent and identically distributed with standard deviation of $1/(\sqrt{2}\,\text{SNR})$. 

We start by using a first-order small-angle approximation of the rotation matrix
\begin{equation}
\label{eq:linInversion}
\begin{split}
\boldsymbol{\Phi} &\approx \mathbf{I}+\gamma\tau\mathbf{A} \\ 
&=
\begin{bmatrix}
 1 & \gamma\tau B_{z} & -\gamma\tau B_{y}\\
 -\gamma\tau B_{z} & 1 & \gamma\tau B_{x}\\
 \gamma\tau B_{y} & -\gamma\tau B_{x} & 1
 \end{bmatrix}\,,
\end{split}
\end{equation} 
where $\mathbf{I}$ is the identity matrix. The magnetic field components can be solved directly and, as each component is measured twice, they can be averaged so that the noise SD in the angular quantity becomes $\sigma_{\gamma\tau \hat{B}_m} = 1/(2\,\mathrm{SNR})$. Here, $m$ is any of the components $x,y$ or $z$, and the noise SD of a magnetic field component can be derived to $\sigma_{\hat{B}_m} = 1/(2 \gamma\tau\ \mathrm{SNR})$.

In reality, the elements of $\boldsymbol{\Phi}$ are estimated with the help of a reference image, which modifies the noise in the elements as derived in the Appendix. %\ref{sec:appendix}. 
Additionally, only two rows of the rotation matrix $\boldsymbol{\Phi}$ can be obtained from the measurements as explained in section~\ref{Sec:State of the art}. Therefore, one row (in our case the first row) has to be derived from the cross product of the adjacent rows, where the cross product contains information about the components of $\boldsymbol{B}$ orthogonal to the direction of $\boldsymbol{B}_{\text{0}}$. These components are no longer subject to independent random noise; consequently, the noise is not reduced by the averaging effect in the linear reconstruction.

So far, the noise analysis was discussed for the reconstruction of the effective $\boldsymbol{B}$-field. 
As mentioned before, in practice, the measurement of $\boldsymbol{B}_{J}$ is contaminated by a background field $\boldsymbol{B}_{\text{B}}$, which must be eliminated by subtracting a second reconstruction. The noise in the two reconstructions is independent, which is why in the case of bipolar reconstruction the noise in the field estimate is reduced by a factor of $\sqrt{2}$ (see Eq.~\eqref{eq:bipolar}). Additionally, as the reference phase $\delta$ is the same for the two data sets, the additional noise due to referencing will cancel in the field subtraction.

In the first-order approximation, we finally obtain for bipolar reconstruction
\begin{equation}
    \sigma_{\hat{B}_y} = \sigma_{\hat{B}_z} \approx \dfrac{1}{2 \gamma\tau\,\mathrm{SNR}}
\end{equation}
and 
\begin{equation}
    \sigma_{\hat{B}_x} \approx \dfrac{1}{2 \sqrt{2} \gamma\tau\,\mathrm{SNR}}\,,
\end{equation}
because $B_x$ is measured twice.

\subsection{Noise analysis of B-field reconstruction: Monte-Carlo simulations}
From the first-order small-angle approximation we can gain intuitive understanding of the statistical uncertainty in the reconstruction of $\boldsymbol{B}_{J}$. However, in reality, the rotation angle $\phi$ can obtain values up to $\pi$ and the linear approximation breaks down.

In order to estimate the influence of noise on the non-linear reconstruction, we carried out a series of Monte-Carlo simulations. Therefore, we generated the last two rows of rotation matrices $\boldsymbol{\Phi}$ for 100 different rotation angles $\phi = \pm \gamma\tau |\boldsymbol{B}|$ taken uniformly between $-\pi<\phi<\pi$,  where the negative angles correspond to $-\boldsymbol{B}$. As before, $\boldsymbol{B}=\boldsymbol{B}_{\text{B}}+\boldsymbol{B}_{J}$, where $\boldsymbol{B}_{J}$ was set to zero and $\phi$ was varied by adjusting $\boldsymbol{B}_{\text{B}}$. The matrices $\boldsymbol{\Phi}$ were generated using the general formula of Rodriguez, as explained in \cite[p.~86--89]{Kraus2014}:
\begin{equation}
    \boldsymbol{\Phi} = e^{\phi \boldsymbol{K}} = \boldsymbol{I} + \sin(\phi)\boldsymbol{K} + (1-\cos(\phi))\boldsymbol{K}^2\,.
\end{equation}
Here, $\boldsymbol{K} = \gamma \tau \boldsymbol{A}/\phi$ is a unitary cross-product matrix associated with the rotation axis. Independent and Gaussian-distributed random noise was generated and superimposed with each element of $\boldsymbol{\Phi}$, according to Eq.~\eqref{eq:SNR}. Subsequently, the first row was derived by the cross product of the other two. The procedure was repeated 100000 times to obtain statistics for the reconstruction quality.

\begin{figure*}[t]
	\centering
	\makebox[1pt]{
	\includegraphics[width=\textwidth]{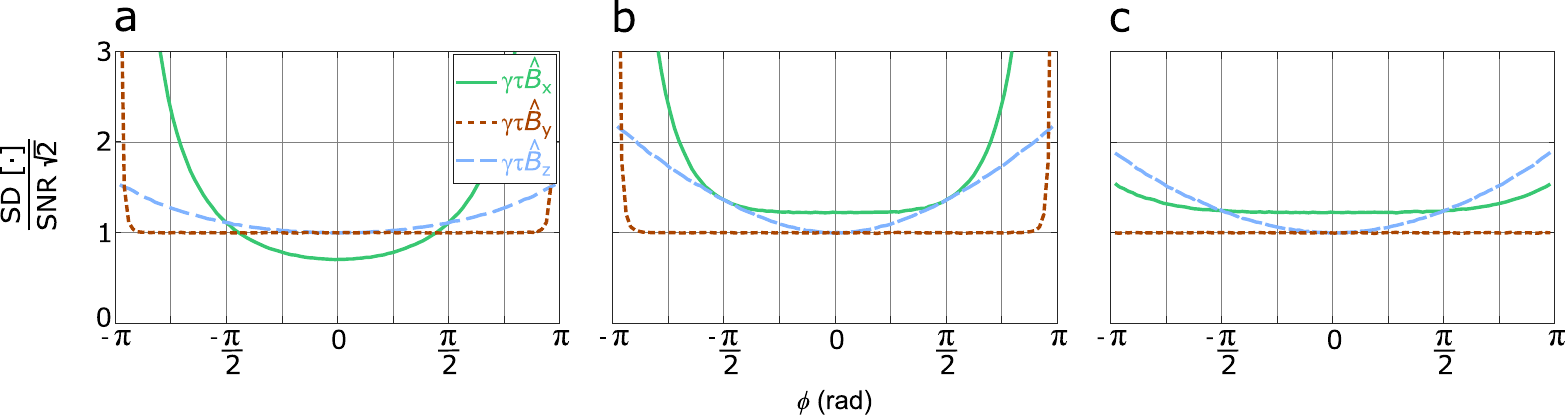}}
	\caption{Single-voxel Monte-Carlo simulations to estimate the influence of noise on three different steps of the non-linear reconstruction as a function of the rotation angle $\phi$. The basis were noisy rotation matrices, where the first row was derived by the cross product of the other two. Displayed are normalized standard deviations of each component of $\boldsymbol{\hat{B}}$, which is the reconstruction of $y$-directional field $\boldsymbol{B}=|\boldsymbol{B}_{\text{B}}|\boldsymbol{e}_{y}$.
	%+\boldsymbol{B}_{J}$, where $\boldsymbol{B}_{J}$ was set to zero.
	$|\boldsymbol{B}_{\text{B}}|$ was adjusted to generate the rotation angles $\phi$ with the negative angles corresponding to the field direction $-\boldsymbol{e}_{y}$. The main field $\boldsymbol{B}_{\text{0}}$ was \textit{x}-directional. The figures show the standard deviations of reconstructions without pre-referencing (a), with pre-referencing before cross product operation (b), and with subsequent orthogonalization using Löwdin's transformation (c).}	
	\label{Fig:SigmaB1}
\end{figure*}

Fig.~\ref{Fig:SigmaB1} illustrates the standard deviation after three intermediate steps of the reconstruction, showcasing their influences on the result. The data are normalized to the input noise $1 /(\sqrt{2}\,\text{SNR})$ corresponding to Eq.\eqref{eq:SNR} without $g_{ij}(\boldsymbol{\Phi})$. 

Fig.~\ref{Fig:SigmaB1}(a) illustrates a case where no referencing with $u$ was applied. Each element of $\boldsymbol{\Phi}$ thus contained the same amount of Gaussian distributed noise. Although this may not be the case in an experimental implementation, one sees that $B_x$ contains $1/\sqrt{2}$ the noise of the other components for small angles of $\phi$, as predicted by the first-order approximation. However, with a rising field strength, \textit{i.e.}, larger rotation angle $\phi$, the noise in this component increases non-linearly and more strongly compared to the components orthogonal to $\boldsymbol{B}_{\text{0}}$. 

The simulations underlying Fig.~\ref{Fig:SigmaB1}(b) include the necessary pre-referencing. For very small angles, the extra phase noise due to the noisy reference phase $\delta$ affects the noise SD only in $\hat{B}_{x}$. Towards larger angles, this effect is visible in $\hat{B}_{z}$. The \textit{y}-component of $\boldsymbol{\hat{B}}$ is not affected, which is in accordance with the analysis presented in the appendix.

\begin{figure*}[t]
	\centering
	\makebox[1pt]{
	\includegraphics[width=\textwidth]{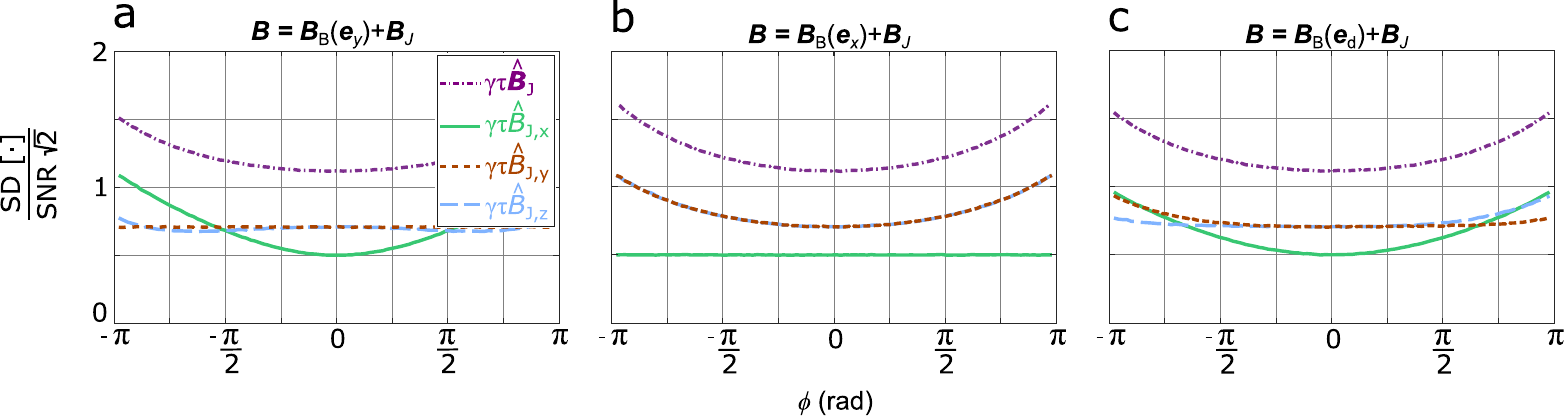}}
	\caption{Single-voxel Monte-Carlo simulations to estimate the standard deviation of each component of $\boldsymbol{\hat{B}}_{J}$ after bipolar reconstruction (Eq.~\eqref{eq:bipolar}), in dependence of the rotation angle $\phi$. In addition, $\sqrt{\operatorname{tr}[\operatorname{cov}(\boldsymbol{\hat{B}}_{J})]}$ (Eq.~\eqref{eq: deviation of the mean}) is presented in purple, dash/dotted lines. $\boldsymbol{B}$ is the effective field $\boldsymbol{B}_{\text{B}}+\boldsymbol{B}_{J}$, where $\boldsymbol{B}_{J}$ was set to zero and $\boldsymbol{B}_{\text{B}}$ was adjusted to generate defined rotation angles $\phi$ with negative angles corresponding to $-\boldsymbol{B}$. The figures represent reconstructions, where $\boldsymbol{B}_{\text{B}}$ was \textit{y}-directional (a), \textit{x}-directional (b), and diagonally oriented in $\boldsymbol{e}_{\text{d}}=[1,1,1]/\sqrt{3}$ (c). The main field $\boldsymbol{B}_{\text{0}}$ was \textit{x}-directional in all cases. }	
	\label{Fig:SigmaB2}
\end{figure*}

Fig.~\ref{Fig:SigmaB1}(c) shows the results after subsequent orthogonalization using the Löwdin transformation. We observe a strong effect towards large angles $\phi$, especially in the \textit{x}-component, which is parallel to $\boldsymbol{B}_{\text{0}}$. 

Fig.~\ref{Fig:SigmaB2} illustrates the standard deviations of the results of a simulated bipolar reconstruction. In comparison to Fig.~\ref{Fig:SigmaB1}, these data sets are arithmetic means of two similar fields (independent noise, identical reference), respectively Eq.~\eqref{eq:bipolar} with $\boldsymbol{B}_{J}=0$. Overall, the noise levels decrease by a factor of $\sqrt{2}$, in comparison to the reconstructions of the effective field $\boldsymbol{B}$ in Fig.~\ref{Fig:SigmaB1}. Further, the additional noise due to the reference phase $\delta$, visible in Fig.~\ref{Fig:SigmaB1}(b--c), was subtracted entirely. Except for very large angles ($\phi>7\pi/8$), the noise SD in each component is lower than $1/(\text{SNR}\sqrt{2})$. Fig.~\ref{Fig:SigmaB2} also shows a measure to assess the expected deviation from the mean of $\boldsymbol{\hat{B}}_J$ (purple line), which can be derived to be the square root of the trace of the covariance matrix:
 \begin{equation}
 \label{eq: deviation of the mean}
 \begin{split}
     \operatorname{SD}[\boldsymbol{\hat{B}}_{J}] 
     &= \sqrt{\operatorname{E}\left[|\boldsymbol{\hat{B}}_J -\operatorname{E}(\boldsymbol{\hat{B}}_J)|^2\right]} \\
     &= \sqrt{\operatorname{tr}\left[\operatorname{cov}(\boldsymbol{\hat{B}}_{J})\right]}\\
     & = \sqrt{\sigma_{B_{J,x}}^{2}+\sigma_{B_{J,y}}^{2}+\sigma_{B_{J,z}}^{2}}\,.
 \end{split}
 \end{equation}

\subsection{Noise analysis of J-field reconstruction}
From the noise in the reconstruction of $\boldsymbol{B}_{J}$, we can also calculate the noise in the current density reconstruction using Eq.~\eqref{eq:Amperes_Law}. For that, we make some simplifications. We assume a constant current density in a homogeneous and isotropic medium. Further, we assume a homogeneous background field that is much larger than $\boldsymbol{B}_{J}$. A simple method for the spatial derivation is to take into account only the two nearest neighbours at $z-l$ and $z+l$
\begin{equation}
\label{eq:deriv1}
\dfrac{d\boldsymbol{\hat{B}}_{J}}{dz}(z) = \dfrac{\boldsymbol{\hat{B}}_{J}(z+l)-\boldsymbol{\hat{B}}_{J}(z-l)}{2l}\,,
\end{equation} 
where $z$ is the coordinate of the voxel in the \textit{z}-direction and $l$ is the voxel sidelength. Assuming equal SNR at $z+l$ and $z-l$, the noise SD of the gradient is approximately $\sigma_{\boldsymbol{G}(z_{n})}=\sigma_{\boldsymbol{\hat{B}}_{J}(z_{n})}/(l\sqrt{2})$. Applying the curl 
\begin{equation}
\label{eq:current_density_diff}
\hat{J}_{x} = \dfrac{1}{\mu_{0}}\left( d\hat{B}_{J,z}/dy-d\hat{B}_{J,y}/dz \right)
\end{equation}
and neglecting the small possible differences in $\sigma_{\hat{B}_{J,z}}$ and $\sigma_{\hat{B}_{J,y}}$, the noise SD of $\hat{J}_{x}$ can be approximated as $\sigma_{\hat{J}_{x}}=\sigma_{\hat{B}_{J,z}}/(l\mu_{0})$.
\subsection{Field reconstruction quality in terms of image SNR}
Using the definition of image SNR in Eq.~\eqref{eq:SNRdef} and the results of the Monte-Carlo simulations, the signal-to-noise ratio of the $\boldsymbol{B}_{J}$ reconstruction ($\text{SNR}[\boldsymbol{\hat{B}_J}]$) can be estimated by 
\begin{equation}
\label{eq:SNR_Bj}
\begin{split}
\text{SNR}[\boldsymbol{\hat{B}}_J] &\overset{\mathrm{def}}{=} \dfrac{|\boldsymbol{\hat{B}}_J|}{\operatorname{SD}[\boldsymbol{\hat{B}}_{J}]}\\ 
&= \dfrac{\gamma\tau|\boldsymbol{\hat{B}}_J|\sqrt{2}}{c}\text{SNR}\,,
\end{split}
\end{equation} 
where $\operatorname{SD}[\boldsymbol{\hat{B}}_{J}]$ is the measure for noise in the vector $\boldsymbol{\hat{B}}_J$ defined in Eq.~\eqref{eq: deviation of the mean}. Further, the scaling factor $c$ depends on the strength and the orientation of $\boldsymbol{B}_{\text{B}}$ and can be read directly from the purple, dash/dotted lines in Fig.~\ref{Fig:SigmaB2}. As $c$ is highest for \textit{x}-directional background fields, a polynomial, normalized to $1/\pi$, was fitted to the data presented in Fig.~\ref{Fig:SigmaB2}(b), to approximate $c$ as a function of $\phi$:
\begin{equation}
\label{eq:polynomial}
% c \approx 0.0018\phi^{4}+0.036\phi^{2}+1.118.
c(\phi) \approx 0.17\left(\dfrac{\phi}{\pi}\right)^{4}+0.35\left(\dfrac{\phi}{\pi}\right)^{2}+1.118\,.
\end{equation}
Note that the results presented in Figs.~\ref{Fig:SigmaB2}(a~and~c) only deviate slightly from Eq.~\eqref{eq:polynomial}.

According to the figure, without any information on the background field, a representative value for the scaling factor would be $c=1.3$. This is close to the worst-case scenario as higher rotation angles may cause phase wrapping. 

To provide a numerical example, let us assume a $|\boldsymbol{B}_{J}| = 10 \text{ nT}$, a homogeneous \textit{x}-directional background field of 60~nT, and a zero-field time $\tau$~=~100~ms, taking into account the $T_2$-relaxation time of grey matter in the $\upmu$T regime of approximately 100~ms. Substituting the rotation angle $\phi=\gamma\tau|\boldsymbol{B}|$ in Eq.~\eqref{eq:polynomial}, $c$ is approximated to be $1.2$. According to Eq.~\eqref{eq:SNR_Bj}, for a required $\text{SNR}[\boldsymbol{\hat{B}}_{J}] >  10$, the voxel SNR needs to be over 32.

The estimation of $\boldsymbol{J}$ using Ampère's law requires the determination of local field gradients, where the noise in the reconstruction is inversely proportional to the voxel side length $l$. This effect should not be underestimated, as the signal strength already scales to the voxel volume $l^3$, the SNR of $\boldsymbol{\hat{J}}$ scales to the fourth power of the voxel sidelength. The quality of the $\boldsymbol{J}$-reconstruction can be determined from the SNR of $\boldsymbol{\hat{B}}_J$, by including the scaling factor $l \mu_0$ in Eq.~\eqref{eq:SNR_Bj}:
\begin{equation}
\label{eq:SNR_J}
\begin{split}
    \text{SNR}[\boldsymbol{\hat{J}}] &\overset{\mathrm{def}}{=} \dfrac{|\boldsymbol{\hat{J}}|}{\operatorname{SD}\boldsymbol{[\hat{J}]}}\\ 
    &\approx \dfrac{\gamma\tau l \mu_0|\boldsymbol{\hat{J}}|\sqrt{2}}{c}\text{SNR}\,.
\end{split}
\end{equation} 
The approximation in Eq.~\eqref{eq:SNR_J} is valid when the voxels involved in the gradient estimation are subject to equal complex voxel SNR. Especially at tissue boundaries, this can cause erroneous assessments due to different relaxation times.

Again, to provide an example, we assume a current density distribution of 0.4~A/m\textsuperscript{2}, a value in accordance with the literature for a stimulation of approximately 4~mA \cite{Neuling2012}. Similar to the example above, $c\approx1.2$ is assumed. If we want to derive $\boldsymbol{\hat{J}}$ with $\text{SNR}[\boldsymbol{\hat{J}}] >  10$ and a voxel-sidelength of 5~mm, a required complex voxel SNR of 130 is estimated. 

\section{Simulated performance of ULF-MRI systems}
\subsection{MRI simulation setup}
\label{sec:mri_setup}
The main factors that determine the SNR profiles of ULF MR images are the sensor arrangement, system noise, and the polarizing field profile. To evaluate the sensitivity of the $\boldsymbol{\hat{B}}_J$ and $\boldsymbol{\hat{J}}$ field reconstruction in a realistic situation, we set up a simulation toolbox incorporating realistic polarizing fields and sensor geometries, as well as time-domain spin evolution based on analytical solutions of Bloch's equation. Assuming ideal gradient fields and instantaneous field switching, gradient-echo sequences can be simulated for arbitrary imaging objects. Both the polarizing field profile and the coupling of the magnetization to the sensor (Eq.~\eqref{eq:Spatial encoding}) were calculated by analytically integrating the Biot--Savart formula over line segments \cite{Hanson2002, zevenhoven2019superconducting}. 

\begin{figure}[t]
    \centering
    \includegraphics[width=0.48\textwidth]{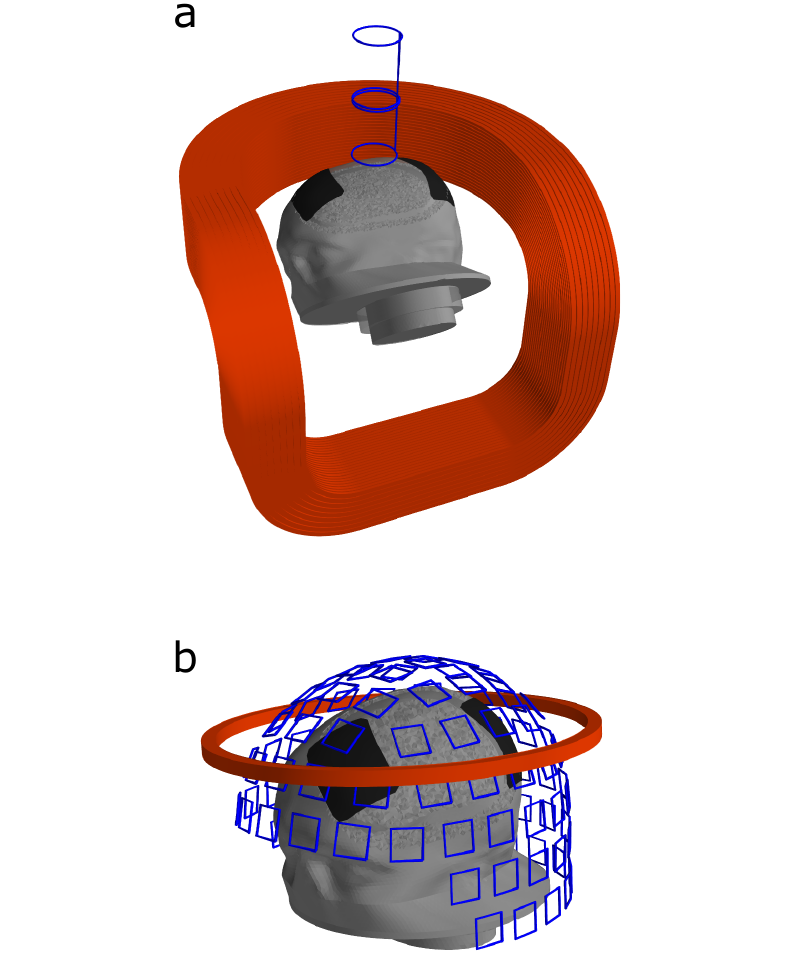}
    \caption{Geometries of the single-channel MRI setup at PTB (a) and the multi-channel MRI setup at Aalto (b). The illustrations include the polarizing coil (red), the receiver coils of the sensors (blue), the head model phantom (gray), and the stimulation electrodes (black).}
    \label{fig:setups}
\end{figure}

\begin{figure}[t]
    \centering
    \includegraphics[width=0.48\textwidth]{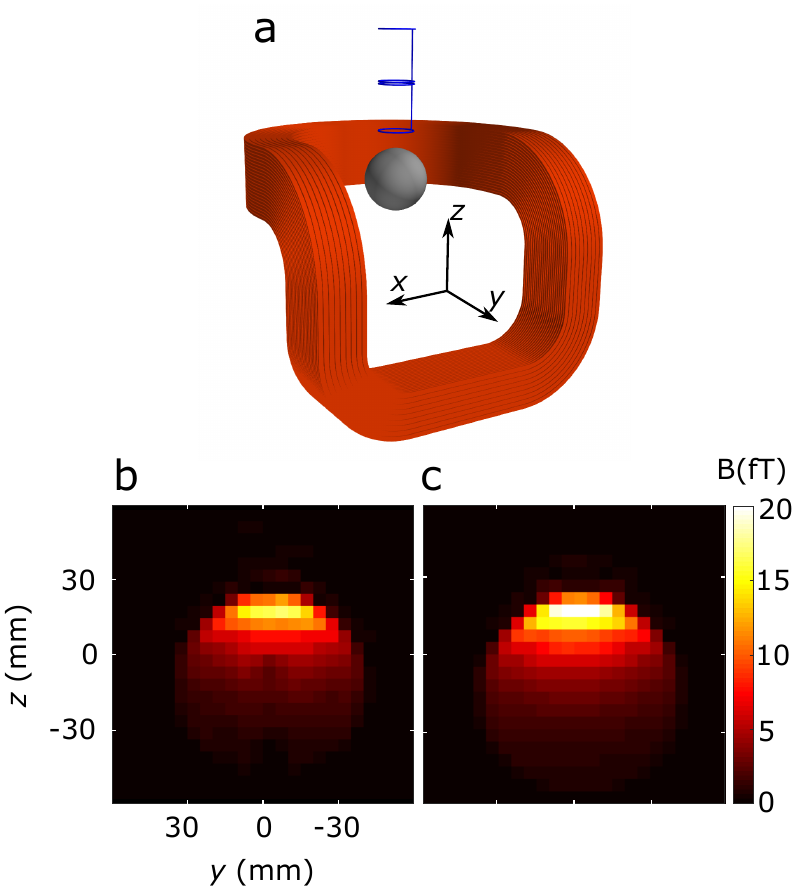}
    \caption{Comparison of measured and simulated MRI images. (a) shows the utilized setup, including the polarizing coil (red), the spherical phantom (gray), and the receiver coil of the sensor (blue). Central slices of reconstructed images, uncorrected for the sensitivity profile, are presented for measurement (b) and simulation (c). Please note that the actual phantom contains a mount for dipolar current electrodes, that is recognisable in the central lower half of the reconstructed measurement, but was not accounted for in the simulations.}
    \label{fig:setup_spherical}
\end{figure}

Two sets of simulations were set up to correspond to the single-channel system with a wire-wound 2\textsuperscript{nd}-order axial gradiometer and a resistive polarizing coil as present at PTB, Berlin, and the multi-channel whole-head system with 102 planar thin-film magnetometers and a compact superconducting polarizing coil built at Aalto University (see Fig.~\ref{fig:setups}). Based on measured values, the sensor noise in the single-channel system was set to 350~$\text{aT/}\sqrt{\text{Hz}}$ and in the multi-channel system to 2~$\text{fT/}\sqrt{\text{Hz}}$. A polarizing current of 50~A was chosen for both setups corresponding to field maximum of 90~mT and mean of 65~mT in the brain compartment for the single-channel system. For the multi-channel system, the field maximum was 115~mT and the mean 70~mT in the brain compartment.

For the evaluation of the simulations, a comparison with actual measurements using the PTB setup was executed. Therefore, a spherical single-compartment phantom (80~mm diameter), filled with an aqueous solution of CuSO\textsubscript{4}+H\textsubscript{2}O to tune the $T_2$-relaxation time to approximately 100~ms, was placed 10~mm below the dewar (nominal warm-cold distance 13~mm). The current in the polarizing coil was set to 20~A, resulting in an inhomogeneous polarizing field of approximately 25~mT. Gradients were set to give a voxel size of $(4.8\times4.8\times4.8)$~mm\textsuperscript{3} and a field of view (FOV) of 115~mm in the phase-encoded directions \textit{y} and \textit{z}. The resulting time signals of the gradient echos were processed to form an array of k-space data. To reduce Gibbs ringing, both the frequency- and the phase-encoding dimensions were tapered with a Tukey window (shape parameter~=~0.5) and the three-dimensional FFT was applied to reconstruct the images. For the simulations, the sphere was approximated by a regular 1-mm spaced grid.

\begin{figure*}[!t]
    \includegraphics[width=1.0\textwidth]{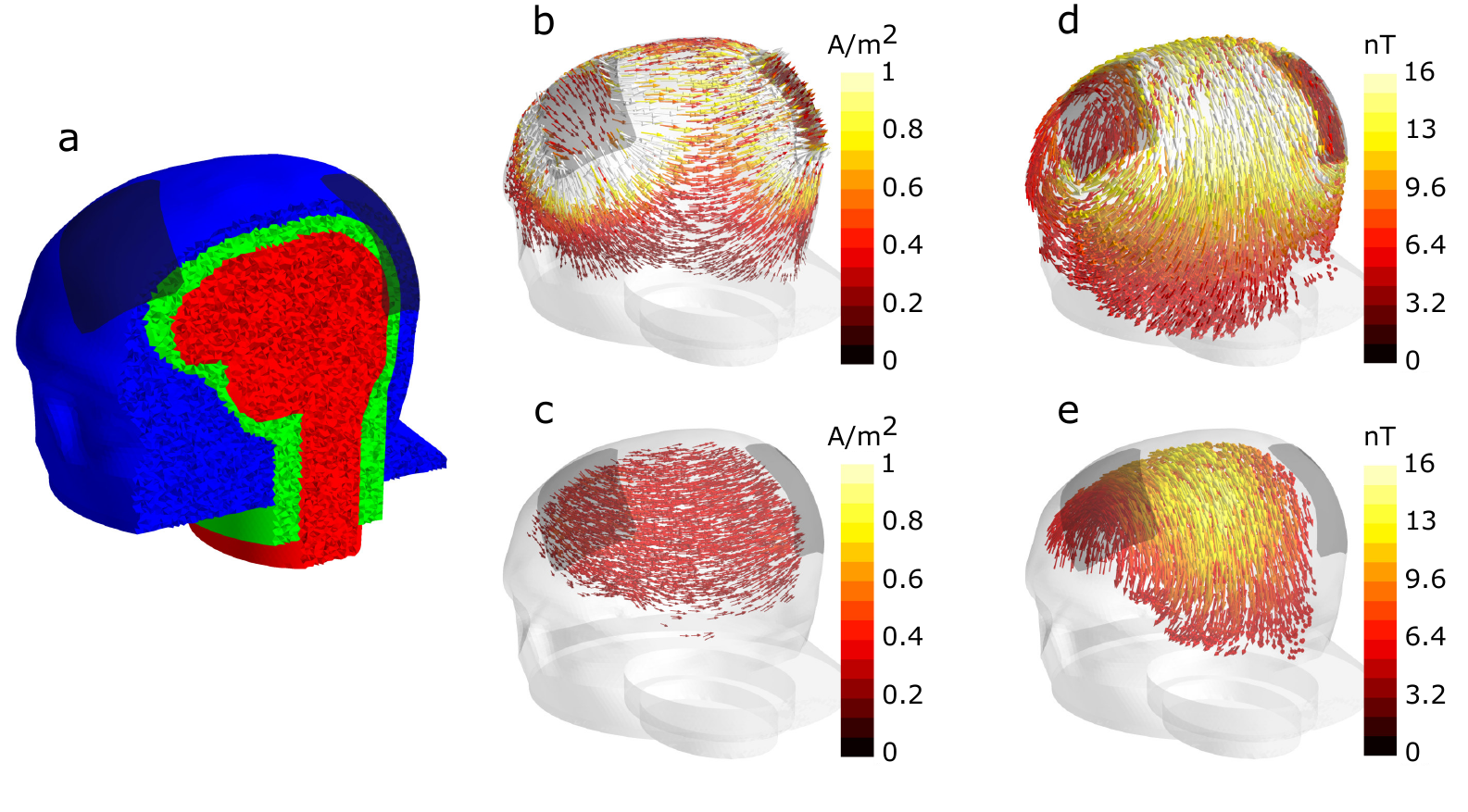}
    \caption{(a) shows the tetrahedral FEM mesh consisting of intra-cranial volume (red), skull (green), and scalp (blue) compartments. The electrodes are illustrated in transparent grey. The simulated current density $\boldsymbol{J}$ is visualized in the scalp (b) and in the brain compartment (c). The simulated magnetic field $\boldsymbol{B}_J$, due to all current flowing in the head, is plotted in the scalp (d) and in the brain (e). The arrow lengths are scaled logarithmically because of the vast magnitude differences especially in the current density. Each subfigure shows only the top 30 (magnitude) percentile of the field in the respective compartment.}
    \label{fig:phantom}
\end{figure*}

Fig.~\ref{fig:setup_spherical} illustrates the setup, accompanied by magnitude images of measurement and simulation. The results reveal a difference in the amplitude of measured and simulated MRI of approximately 25\%, probably subject to multiple origins. A shielding coil reduces the polarizing field of the actual setup, which was not accounted for in the simulations. Also, winding errors due to the relatively complex geometry of the polarizing coil reduce the current--field ratio. In addition, the true warm--cold distance of the dewar could vary depending on the helium level and the phantom mount also might have inaccuracy in the mm range. Taking all these uncertainties into account, the simulated MRI sequence resembles the realistic conditions found in actual measurements.

\subsection{MRI simulations with head model}
In the next step, the simulation setup was used to generate full CDI sequences with the single-channel system, as well as the multi-channel system, using the $\boldsymbol{B}_{J}$ distribution derived from finite-element-method (FEM) simulations of a realistic head model. This model is based on CT scans of a human head \cite{Hunold2019} and contains three compartments as shown in Fig.~\ref{fig:phantom}(a). The conductivity in the outermost scalp compartment was set to 0.22~S/m, in the skull compartment to 0.01~S/m, and in the innermost brain compartment to 0.33~S/m. The two stimulation electrodes were positioned roughly 10~cm apart, one on the forehead and the other one on the side of the head, as shown in Fig.~\ref{fig:phantom}. The conductivity of the electrodes was set to 1.4~S/m.

The FEM simulations to obtain the current density $\boldsymbol{J}$ and the resulting magnetic field $\boldsymbol{B}_{J}$ were conducted in the Comsol Multiphysics software based on the generalized minimal residual method (GMRES). Current flow was realized by setting zero potential on the outer surface of the cathode and applying a total current of 4.5~mA to the outer surface of the anode. For the calculation of $\boldsymbol{B}_{J}$, a spherical air compartment (2~m in diameter) was added to the model, ensuring a negligible effect of the magnetic isolation boundary condition. 

Patterns of the simulated current density and the associated magnetic field are shown in Fig.~\ref{fig:phantom}. Due to the low conductivity of the skull, most of the current flows in the scalp compartment. In the vicinity of the electrode boundary, $|\boldsymbol{J}|$ was up to 15~A/m$^2$. The maximal current density in the brain compartment below the electrodes was about 0.5~A/m$^2$. In relation to that, the magnetic field appeared smoother, yielding maximal field strengths of 20~nT in the scalp and 12~nT in the brain compartment. The maximum of the field magnitude in the brain compartment is localized in between the electrodes, just beneath skull layer. In contrast, the maximal current density in the brain is localized beneath the electrodes.

\begin{figure}[!t]
    \centering
    \includegraphics[width=0.48\textwidth]{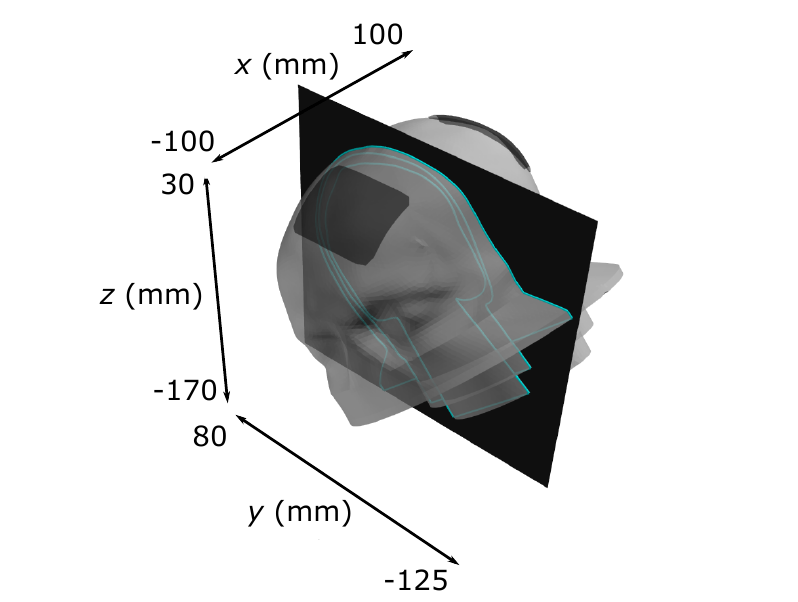}
    \caption{The slice between the electrodes shown in Figs.~\ref{fig:field_comparison} and \ref{fig:system_comparison}. Note that images are shown from the back of the head.}
    \label{fig:slice}
\end{figure}

For the MRI simulations, the head model was positioned in the FOV of the two described scanner arrangements, similar to how the positioning of a head would be in an actual measurement setup (compare with Fig.~\ref{fig:setups}). The scalp-sensor distance was 16~mm for the single-channel setup and 20--35~mm for the multi-channel setup, taking into account the individual warm-cold distances of the two systems plus 3~mm to compensate for the amplitude differences found in the comparison with actual measurements, as described in Sec.~\ref{sec:mri_setup}. The magnetization was discretized to tetrahedral elements derived from the geometry of the Comsol model. The time evolution of the magnetic moment was simulated for the center of each element. The $T_2$-relaxation time for the brain compartment was set to 106~ms and for the scalp compartment to 120~ms \cite{Zotev2009}. For simplicity, as the spin density in the skull is insignificant compared to soft tissue, this compartment was assumed to have no magnetization at all. The average tetrahedron sidelengths were approximately 3.5~mm in the brain and 2.5~mm in the scalp. Gradients were set to give a voxel size of $(5\times5\times5)$~mm\textsuperscript{3} and a field of view (FOV) of 220~mm in the phase-encoded directions. As performed for the spherical phantom, both the frequency- and phase-encoding dimensions were tapered with a Tukey window (shape parameter\,=\,0.5) before computing the three-dimensional FFT. For the multi-channel system, images of each sensor were combined voxel-wise using the coupling field information as described in \cite{zevenhoven2019superconducting}.

\begin{figure}[!t]
    \centering
    \includegraphics[width=\columnwidth]{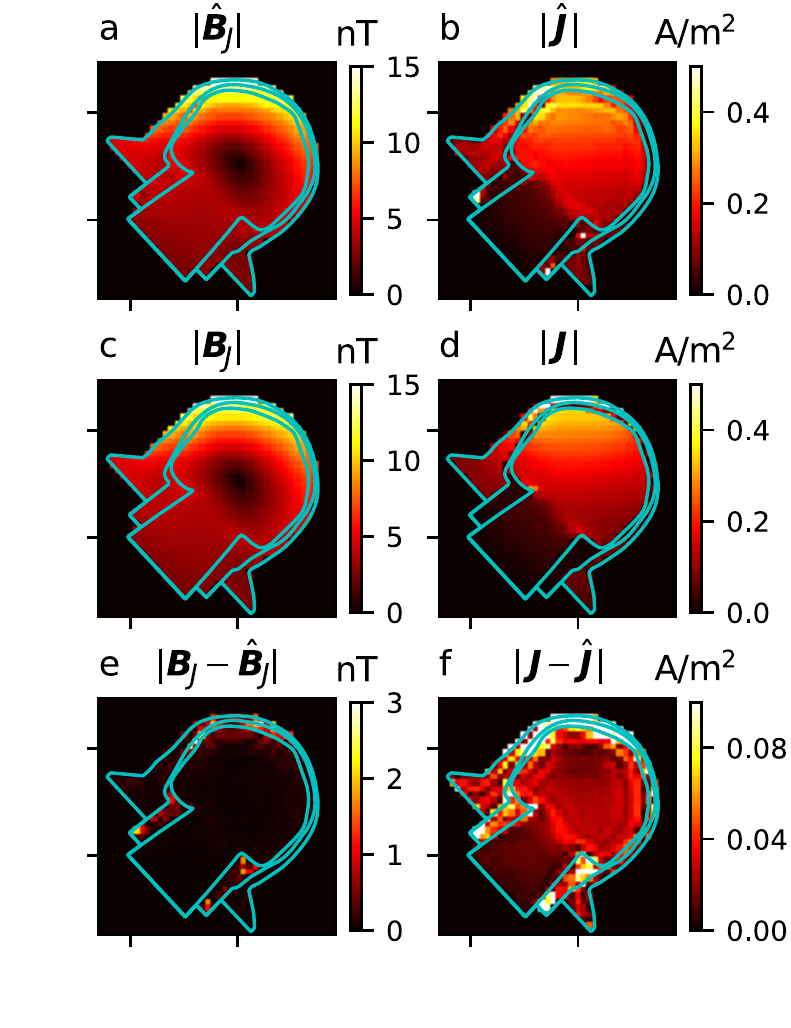}
    \caption{Comparison of the simulated noiseless CDI reconstructions and the FEM solutions. (a) shows the reconstructed magnetic field and (b) the reconstructed current density. (c) and (d) show the respective FEM solutions, and (e) and (f) display the absolute differences between reconstructions and the FEM solutions. The FEM fields are linearly interpolated from the FEM nodal values to the $(5\times5\times5)$~mm\textsuperscript{3} voxel grid. The reconstructions are masked to zero outside the head model. Note that the color axes of the bottom-most figures differ from the others by a factor of 5.}
    \label{fig:field_comparison}
\end{figure}

%\begin{figure}
%    \centering
%    \raisebox{-0.5\height}{\includegraphics[width=0.48\textwidth]{snr_per_amperes_aalto.pdf}}
%    \raisebox{-0.5\height}{ \includegraphics[width=0.48\textwidth]{phantom_slices.png}}
 %   \caption{SNR of $B_J$ per current (Amperes)  during the polarization period.}
  %  \label{fig:field_comparison}
%\end{figure}

\begin{figure}[t!]
    \centering
    \includegraphics[width=\columnwidth]{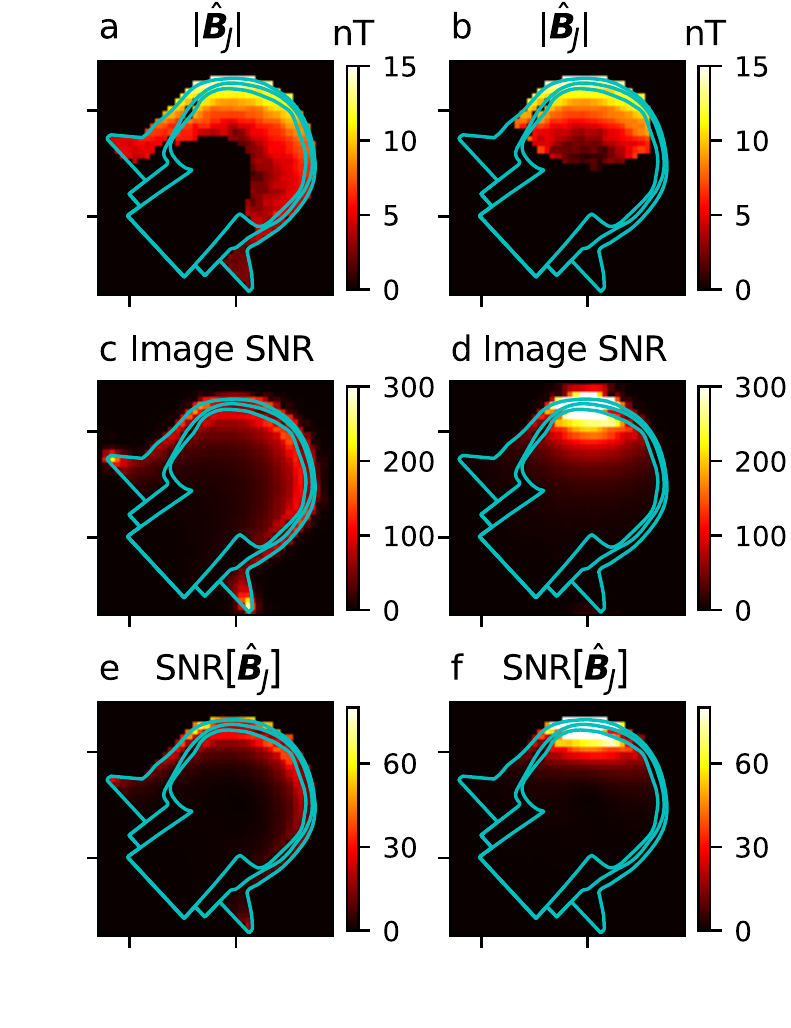}
    \caption{Comparison of system performances of the Aalto multi-channel (a, c, e) and the PTB single-channel (b, d, f) ULF-MRI setups. (a) and (b) show $|\boldsymbol{\hat{B}}_J|$ reconstructions of CDI simulations thresholded above SNR$=20$ and inside the head model. (c) and (d) show the SNR maps of the simulated magnitude images and (e) and (f) contain estimates of SNRs of $\boldsymbol{B}_{J}$ in both systems.}
    \label{fig:system_comparison}
\end{figure}

\subsection{Results}
Fig.~\ref{fig:field_comparison} shows a comparison between the FEM solutions $|\boldsymbol{B}_{J}|$ and $|\boldsymbol{J}|$, and the corresponding field reconstructions $|\boldsymbol{\hat{B}}_{J}|$ and $|\boldsymbol{\hat{J}}|$ of the simulated zero-field sequence without any noise. The reconstructed magnetic field $|\boldsymbol{\hat{B}}_{J}|$ resembles closely the FEM solution $|\boldsymbol{B}_{J}|$, which was used as an input to the MR simulations. Notable differences are found inside the skull, which is expected due to the lack of magnetization, as well as on the top parts of the scalp at the field maximum. The difference image reveals ringing artifacts in the intra-cranial volume, leading to error fields up to approximately 1 nT. 

The difference between the reconstructed current density $|\boldsymbol{\hat{J}}|$ and the corresponding FEM solution $|\boldsymbol{J}|$ is more prominent. Although no noise was added to the simulated data, errors in the finite-difference approximations and artifacts in $\boldsymbol{\hat{B}}$ add up, so that the field estimate near the skull is highly distorted. The intra-cranial fields show greater resemblance, although a notable ringing-artifact from the skull can be seen in $|\boldsymbol{\hat{J}}|$. 

Fig.~\ref{fig:system_comparison} displays the performance of the two ULF-MRI setups with the simulated imaging sequence described in Sec.~\ref{sec:mri_setup}. Figs.~\ref{fig:system_comparison}(a--b) show field reconstruction magnitude $|\boldsymbol{\hat{B}}_J|$ for a CDI sequence with 50~A polarizing current. The time-domain echo signals were superimposed with Gaussian noise of 2~$\text{fT/}\sqrt{\text{Hz}}$ and 0.35~$\text{fT/}\sqrt{\text{Hz}}$ for the multi-channel system and the single-channel system, respectively. The reconstruction quality is highly dependent on the SNR of the underlying ULF MR images, which is shown in Figs.~\ref{fig:system_comparison}(c--d). With the ultra-sensitive single-channel setup, one achieves sensitivity in depth to the intra-cranial volume whereas the multi-channel setup gives a broader sensitivity pattern on the scalp and directly under the skull. Figs.~\ref{fig:system_comparison}(e--f) illustrate estimates of the SNR maps of $\boldsymbol{\hat{B}}_J$, corresponding to the images in Figs.~\ref{fig:system_comparison}(a--b). The maps are derived from the noiseless $\boldsymbol{\hat{B}}_J$ and the SNR maps using Eq.~\eqref{eq:SNR_Bj} with $c=1.3$.

\section{Discussion}
Hömmen \textit{et al.}~\cite{Hoemmen2019} concluded that an increase in image SNR of their setup is necessary for a successful \textit{in-vivo} implementation of current-density imaging. However, based on measurements using simple phantoms, no exact numbers for the requirements in terms of SNR could be presented.  

This work provides a profound understanding of the influence of noise on the reconstruction of the magnetic field $\boldsymbol{B}_{J}$ and the current density $\boldsymbol{J}$. The linearization of the field reconstruction gives an approximate relationship between the image SNR and the statistical uncertainty in the field estimates. Further, Monte-Carlo simulations were used to derive the statistical uncertainty in the presence of large background fields where the non-linearities take effect. The presented link between image SNR and noise in the reconstruction allows the determination of the necessary SNR for the estimates $\boldsymbol{\hat{B}}_{J}$ and $\boldsymbol{\hat{J}}$ within a predefined uncertainty. It also enables the assessment of the performance of specific ULF-MRI systems for zero-field-encoded CDI directly from acquired or simulated image data.

In order to retain constant image SNR in the Monte-Carlo simulations, we adjusted $|\boldsymbol{B}_\text{B}|$ to vary $\phi=\gamma\tau|\boldsymbol{B}_\text{B}|$. We set the zero-field-encoding time to $\tau=T_{2}$, which yields maximum $\text{SNR}[\boldsymbol{B}_{J}]$ according to \cite{Vesanen2014}. However, the non-linear dependence of $\text{SNR}[\boldsymbol{B}_{J}]$ on $\phi$ suggests that there is an optimum set of parameters for each specific case. In reality, the effective background field will be roughly constant over the measurement periods and $\tau$ should be adjusted to obtain maximum $\text{SNR}[\boldsymbol{B}_{J}]$. If the relaxation times are known, Eq.~\eqref{eq:SNR_Bj} and Eq.~\eqref{eq:polynomial} can be utilized to create a cost function that provides parameters for maximum reconstruction quality. It should be mentioned that the optima for $\tau$ are flat and close to $T_{2}$ for small background fields. An adjustment of $\tau$ seems worth in the case of very large background fields, where up to 12\% can be gained in  $\text{SNR}[\boldsymbol{\hat{B}}_{J}]$ compared to $\tau=T_2$. Furthermore, it should be kept in mind that $\phi<\pi$ should be fulfilled to prevent ambiguity in the field reconstruction.

To analyze their performance and suitability for \textit{in-vivo} CDI, our two ULF MRI systems were examined in realistic image simulations. One was the system of Hömmen \textit{et al.}, including an optimized polarizing setup, and the second was a whole-head multi-channel system built at Aalto University. Key features that determine the SNR, such as the polarizing-field pattern, the coupling profile to the sensor, and noise, were accurately modeled. The estimates of $\boldsymbol{B}_{J}$ and $\boldsymbol{J}$ were derived from FEM simulations using a three-compartment head model. The peak current densities in intra-cranial tissue are similar to literature values, when scaled to the applied current of 4.5~mA \cite{Miranda2006, Neuling2012}. However, the three-compartment model neglects the fact that current is partly shunted by cerebrospinal fluid (CSF), which has a higher conductivity compared to grey- and white-matter tissue \cite{Salvador2010, Neuling2012}. 

The $\boldsymbol{B}_{J}$-field distribution served as input for MRI simulations, emulating the entire sequence. Taking into account the insights from the Monte-Carlo simulations and the calculated SNR of the single-channel setup, the required improvement in SNR compared to \cite{Hoemmen2019} can now be specified. The simulations verify that the optimized polarization profile is sufficient. The peak SNR of the multi-channel setup is lower compared to the single-channel setup due to a higher sensor noise and different field coupling. A broader sample coverage, on the other hand, is provided by the multi-channel setup. The comparison between the two systems revealed that both high sensitivity and large sample coverage are required for current-density imaging usable for conductivity estimation.

It should be mentioned that both systems were evaluated with 50-A polarizing current, which represents a close to maximum level for the room-temperature coil used with the single-channel device, whereas the superconducting polarizing coil used with the multi-channel device might be able to carry 2--4 times more current. Such an increase in the polarizing current benefits the image SNR and the SNR of the field estimates by the same factor. However, approaching such high fields will cause flux trapping in the sensor\cite{luomahaara2011, AlDabbagh2018, luomahaara2018} and the superconducting filaments of the coil \cite{Zevenhoven2011MSc, Lehto2017}, which has to be dealt with. Also larger currents required for the compensation of the field transient \cite{nieminen2011} can cause excessive heating in the compensation coils, requiring more sophisticated techniques \cite{zevenhoven2015}.

Besides noise, spatial leakage from the FFT has a significant influence on the quality of the reconstruction. Appropriate windowing of the k-space data manipulates the spatial response function of the voxels, effectively reducing the far-reaching leakage at the cost of a smoothed resolution. However, with the applied imaging and reconstruction procedures, leakage artifacts could not be entirely eliminated, yielding noticeable reconstruction errors, especially visible in the $\boldsymbol{\hat{J}}$-field. Besides spatial filtering, an effective method to reduce ringing artifacts in MRI is to apply more k-steps. However, this might not be applicable to \textit{in-vivo} CDI as it would increase the measurement time significantly. Additionally, post-processing methods, for example ``total variation constrained data extrapolation'' \cite{Block2008}, might reduce the artifacts without decreasing the image resolution. 

The $\boldsymbol{J}$ reconstructions of both systems show limitations in thin tissue structures like the scalp. This is most probably due to the chosen resolution of $5 \times 5 \times 5 \text{ mm}^{3}$, which does not allow sufficient gradient calculations in these areas. Reducing the voxel size to 1--2-mm resolution would increase the quality of the $\boldsymbol{\hat{J}}$-fields, but again at the cost of longer overall measurement time and lower SNR. Generally, the simulations show that the $\boldsymbol{B}_{J}$ reconstruction is more reliable than the $\boldsymbol{J}$ reconstruction, as artifacts strongly affect the gradient estimation.

Shall the reconstructions be used to fit individual conductivity values, superior results are expected when the $\boldsymbol{\hat{B}}_{J}$-field 
is used as the measurement data. However, magnetic fields arising from the current leads should be either modeled or eliminated from the data. One way to exclude these fields would be to consider only closed path integrals of $\boldsymbol{\hat{B}}_{J}$ and to apply the integral form of Ampère's law. It remains to be answered whether this only enables to derive bulk conductivity values only, rather than spatially resolved conductivity mapping. Methods for this have not been presented so far and should be subject to further research. 

\section{Conclusion}
 We introduced methods to gain quantitative information about the effect of stochastic uncertainty on the non-linear reconstruction in zero-field-encoded current-density imaging (CDI). The work provides means to determine the ability of specific ultra-low-field MRI setups to reach acceptable signal-to-noise ratios in field reconstructions based on image SNR and to assess necessary improvements in, \textit{e.g.}, noise performance or polarizing field strength. By simulations, we evaluated the reconstruction quality of two existing setups under realistic conditions. We showed that current technology in ULF MRI is suitable for \textit{in-vivo} CDI in terms of SNR. In addition, we encountered reconstruction errors due to a limited resolution and image artifacts requiring further research and development of more accurate reconstruction techniques.
 
%\section*{Data availability statement}
%--- The data statement will be generated during submission, but this is what we aim for ---\\
%The simulation data generated for this study are available on request to the corresponding authors.
 
\section*{Author contributions}
PH, AM, and RK contributed to the conception of the study. PH and AM performed the simulations, analyzed the results, wrote the first draft of the manuscript and revised the manuscript based on the annotations of the co-authors. PH performed comparative measurements. AM contributed to the theory part in the Appendix. AH, RM, and JH developed the head model and performed FEM simulations. All the authors contributed to manuscript revision and read and approved the submitted version. The study was supervised by JH, KZ, RI, and RK.
 
\section*{Funding}
This project has received funding from the European Union’s Horizon 2020 research and innovation programme under grant agreement No~686865. It was partly supported by Vilho, Yrjö and Kalle Väisälä Foundation, by Project 2017~VF~0035 of the Free State of Thuringia, and by DFG Ha~2899/26-1.

\section*{Acknowledgement}
The authors thank Jan-Hendrik Storm for fruitful discussions on the FEM simulations. 

\bibliographystyle{frontiersinHLTH&FPHY} % for Health, Physics and Mathematics articles
\bibliography{references}

%\bibliography{references} 
%\bibliographystyle{ieeetr}

\appendix
\section{Noise in the rotation matrix estimate}
\label{sec:appendix}
Let us define a vector of the real parts of zero-field-encoded raw voxel data as $\boldsymbol{a} = [\operatorname{Re}[v_x], \operatorname{Re}[v_y],\operatorname{Re}[v_z]]^\top$ and of the imaginary parts as $\boldsymbol{b} = [\operatorname{Im}[v_x], \operatorname{Im}[v_y], \operatorname{Im}[v_z]]^\top$, where $v_x$, $v_y$ , and $v_z$ are the voxel values for $x$-, $y$-, and $z$-directional starting magnetization. Now the whole data can be expressed as a combination of these as $\boldsymbol{a}+i\boldsymbol{b}$ which is a 3D vector with complex elements. Applying the same phase correction obtained from a reference image for each element, we can write the data as $(\boldsymbol{a}+i\boldsymbol{b})e^{-i\delta}$. Note that applying the phase correction does not change the noise distribution in the real and imaginary parts when they both have equal amount of noise.

The phase-corrected zero-field-encoded data can be converted to a rotation matrix by normalizing the rows corresponding to $\boldsymbol{a}$ and $\boldsymbol{b}$. Apart from noise, vectors $\boldsymbol{a}$ and $\boldsymbol{b}$ should have the same norm, which lets us approximate $|\boldsymbol{a}| \approx |\boldsymbol{b}| \approx |u_0|$, where $u_0$ is the voxel value in a (ideal) reference image. Thus, we can normalize the imaginary and real parts with a common factor $(\boldsymbol{a}+i\boldsymbol{b})e^{-i\delta}/|u_0|$. The elements of third and second row of the rotation matrix can now be read from the real and imaginary parts of this quantity and the first row 
%$1/|u_0|(\boldsymbol{a}+i\boldsymbol{b})e^{i\delta}$
could be derived as a cross product of the normalized rows.

To derive the effect of noise in the matrix elements, we assume that $v_x$, $v_y$, and $v_z$ are all contaminated by complex Gaussian noise. If the normalization and reference phase were noise-free, the standard deviation of the elements of the rotation matrix would read
\begin{equation*}
    \sigma_{\boldsymbol{\Phi},0} = \frac{1}{\sqrt{2}\,\mathrm{SNR}}\,,
\end{equation*}
where SNR is defined as in \eqref{eq:SNRdef}. Since the noise in the elements is uncorrelated we can write the noise covariance matrix for the both rows as $\sigma_{\boldsymbol{\Phi},0}^2\boldsymbol{I}$.

Next, let us study how the noise in the reference phase and normalization affect the noise covariance matrix of the third row corresponding to $\boldsymbol{a}$ when $\delta=0$. Adding noisy reference phase increases noise in $\boldsymbol{a}$ in the direction of $\boldsymbol{b}$ as it can be considered as a small rotation in the plane spanned by $\boldsymbol{a}$ and $\boldsymbol{b}$. In the limit of high SNR, the standard deviation of phase noise $\Delta\delta$ is approximately $1/(\sqrt{2}\,\mathrm{SNR})$. The noise in the phase factor can be expanded as $e^{i\Delta\delta} \approx 1 + i\Delta\delta$, and in the limit of small perturbation in the noise in $\boldsymbol{a}$ would be $\Delta\delta\boldsymbol{b}$, giving $|\boldsymbol{b}|^2/(\sqrt{2}\,\mathrm{SNR})^2 \approx |u_0|^2/(\sqrt{2}\,\mathrm{SNR})^2$ extra variance in the direction of $\boldsymbol{b}$. The noise covariance of the non-normalized but phase-reference-noise-affected third row would be approximately $|u_0|^2\sigma_{\boldsymbol{\Phi},0}^2(\boldsymbol{I} + \boldsymbol{P_b})$, where $\boldsymbol{P}_{\boldsymbol{b}} = \boldsymbol{b}_0\boldsymbol{b}_0^{\top}/|\boldsymbol{b}_0|^2$, where $\boldsymbol{b}_0$ is the expected value of the vector $\boldsymbol{b}$.

The effect of noise in the normalization constant can be analyzed by interpreting the noise $\boldsymbol{\epsilon}$ in the vector elements of $\boldsymbol{a} = \operatorname{E}[\boldsymbol{a}] + \boldsymbol{\epsilon} = \boldsymbol{a}_0 + \boldsymbol{\epsilon}$ also as a small perturbation
\begin{equation*}
    \frac{1}{|\boldsymbol{a}|} \approx \frac{1}{|\boldsymbol{a}_0|}\left(1- \frac{\operatorname{\boldsymbol{a}_0^\top\boldsymbol{\epsilon}}}{|\boldsymbol{a}_0|^2}\right) ,
\end{equation*}
where only the first-order term in $\boldsymbol{\epsilon}$ is considered. In this case, $\boldsymbol{\epsilon}$ will also contain the effect of noisy phase referencing. Normalizing the third row then gives approximately
\begin{equation*}
\begin{split}
    \frac{\boldsymbol{a}}{|\boldsymbol{a}|} &\approx \frac{\boldsymbol{a}_0}{|\boldsymbol{a}_0|} + \frac{\boldsymbol{\epsilon}}{|\boldsymbol{a}_0|} -  \frac{\operatorname{\boldsymbol{a}_0^\top\boldsymbol{\epsilon}}}{|\boldsymbol{a}_0|^2} \frac{\boldsymbol{a}_0}{|\boldsymbol{a}_0|} \\ \\
    & =
    \frac{\boldsymbol{a}_0}{|\boldsymbol{a}_0|} + \left(\boldsymbol{I}-\frac{\boldsymbol{a}_0\boldsymbol{a}_0^{\top}}{|\boldsymbol{a}_0|^2}\right)\frac{\boldsymbol{\epsilon}}{|\boldsymbol{a}_0|} \\ \\
    & = \frac{\boldsymbol{a}_0}{|\boldsymbol{a}_0|} + \left(\boldsymbol{I}-\boldsymbol{P}_{\boldsymbol{a}}\right)\frac{\boldsymbol{\epsilon}}{|\boldsymbol{a}_0|},
\end{split}
\end{equation*}
where the matrix $\boldsymbol{I} - \boldsymbol{P}_{\boldsymbol{a}} =  \boldsymbol{I}-\boldsymbol{a}_0\boldsymbol{a}_0^{\top}/|\boldsymbol{a}_0|^2$ projects out any component in direction of $\boldsymbol{a}_0$, \textit{i.e.}, $(\boldsymbol{I} - \boldsymbol{P}_{\boldsymbol{a}})\boldsymbol{a}_0=0$, but leaves components orthogonal to $\boldsymbol{a}_0$ unaffected. The noise is thus the same as in the case of noiseless normalization but the noise contribution in the direction of $\boldsymbol{a}_0$ is cancelled. For example, when $\boldsymbol{a}$ is roughly $y$ directional, the noise in the direction of $y$ is removed by the normalization. 

Normalizing the row exactly to unit norm modifies the noise approximately with the linear transformation $|\boldsymbol{a}_0|^{-1}(\boldsymbol{I} - \boldsymbol{P}_{\boldsymbol{a}}) \approx |u_0|^{-1}(\boldsymbol{I} - \boldsymbol{P}_{\boldsymbol{a}})$ giving a new covariance matrix $$(\boldsymbol{I} - \boldsymbol{P}_{\boldsymbol{a}}) \sigma_{\boldsymbol{\Phi},0}^2(\boldsymbol{I} +  \boldsymbol{P}_{\boldsymbol{b}})(\boldsymbol{I} - \boldsymbol{P}_{\boldsymbol{a}})
= \sigma_{\boldsymbol{\Phi},0}^2(\boldsymbol{I} - \boldsymbol{P}_{\boldsymbol{a}} + \boldsymbol{P}_{\boldsymbol{b}})$$
as $\boldsymbol{P}_{\boldsymbol{a}} \boldsymbol{P}_{\boldsymbol{b}} = 0$ because $\boldsymbol{a}_0$ and $\boldsymbol{b_0}$ are orthogonal and $\boldsymbol{P}_{\boldsymbol{a}}^2 =
\boldsymbol{P}_{\boldsymbol{a}}$ because the operator is a projection. 

The noise covariance can be diagonalized in the row basis of the (noiseless) rotation matrix $\boldsymbol{\Phi}_0$, \textit{i.e.},
$$\sigma_{\boldsymbol{\Phi},0}^2(\boldsymbol{I} - \boldsymbol{P}_{\boldsymbol{a}} + \boldsymbol{P}_{\boldsymbol{b}}) = \boldsymbol{\Phi}_0^\top 
\begin{bmatrix}
\sigma_{\boldsymbol{\Phi},0}^2 & 0 & 0 \\
0 & 2\sigma_{\boldsymbol{\Phi},0}^2 & 0 \\
0 & 0 & 0 
\end{bmatrix} \boldsymbol{\Phi}_0, $$
\textit{i.e.}, there is zero variance in the direction of $\boldsymbol{a_0}$, double variance in the direction of $\boldsymbol{b_0}$ and non-modified variance in the direction of the first row. 

Similar analysis can be made for the second row of the rotation matrix estimate giving the following noise covariance
$$\sigma_{\boldsymbol{\Phi},0}^2(\boldsymbol{I} + \boldsymbol{P}_{\boldsymbol{a}} - \boldsymbol{P}_{\boldsymbol{b}}) = \boldsymbol{\Phi}_0^\top 
\begin{bmatrix}
\sigma_{\boldsymbol{\Phi},0}^2 & 0 & 0 \\
0 & 0 & 0 \\
0 & 0 & 2\sigma_{\boldsymbol{\Phi},0}^2 
\end{bmatrix} \boldsymbol{\Phi}_0.$$

The analysis in this section considers only the noise in the estimate of the rotation matrix. In the small angle approximation in Sec.~\ref{sec:linear_approx}, we can use this information directly to explain the noise behaviour of the magnetic field estimates. However, when the rotation angle $\phi$ increases, non-linear reconstruction has to be applied yielding effects that we study using Monte-Carlo simulations.

For small rotation angles, the noise in $B_y$ and $B_z$ is explained by the noise variance in the rotation matrix. To explain the noise in the estimates of the effective magnetic field component $B_x$, we still have to take in to account that, due to the same reference phase noise, the noise in $\Phi_{3,2}$ and $\Phi_{2,3}$ is correlated. The covariance between the elements can be derived to be $-\sigma_{\boldsymbol{\Phi},0}^2$. In consequence, the variance of $\gamma\tau B_x \approx (\Phi_{3,2} -\Phi_{2,3})/2$ becomes
\begin{equation*}
\begin{split}
\operatorname{Var}[\gamma\tau B_x]=& \frac{\operatorname{Var}[\Phi_{2,3}]}{4}+\frac{\operatorname{Var}[\Phi_{3,2}]}{4}\\ 
&- 2 \frac{ \operatorname{Cov}[\Phi_{3,2},\Phi_{2,3}]}{4}
\end{split}
\end{equation*}
which results in $3\sigma_{\boldsymbol{\Phi},0}^2/2$ corresponding to the Monte-Carlo estimate in Fig.~\ref{Fig:SigmaB1}(b). This analysis considers a single reconstruction of effective $B_x$. In bipolar reconstruction, as presented in Eq.~\eqref{eq:bipolar}, the additional noise due to the noisy phase reference cancels out.

%Well, what should come out is something like this:

%$$\sigma_{\boldsymbol{\Phi},0}^2(\boldsymbol{I} - \boldsymbol{P}_{\boldsymbol{a}} + \boldsymbol{P}_{\boldsymbol{b}}) = \boldsymbol{\Phi}_0^\top 
%\begin{bmatrix}
%\sigma_{\boldsymbol{\Phi},0}^2 & 0 & 0 \\
%0 & \sigma_{\boldsymbol{\Phi},0}^2+(2\sigma_{\boldsymbol{\Phi},0})^2 & 0 \\
%0 & 0 & 0 
%\end{bmatrix} \boldsymbol{\Phi}_0$$,

%$$\sigma_{\boldsymbol{\Phi},0}^2(\boldsymbol{I} + \boldsymbol{P}_{\boldsymbol{a}} - \boldsymbol{P}_{\boldsymbol{b}}) = \boldsymbol{\Phi}_0^\top 
%\begin{bmatrix}
%\sigma_{\boldsymbol{\Phi},0}^2 & 0 & 0 \\
%0 & 0 & 0 \\
%0 & 0 & \sigma_{\boldsymbol{\Phi},0}^2+(2\sigma_{\boldsymbol{\Phi},0})^2 
%\end{bmatrix} \boldsymbol{\Phi}_0$$.

%It matches the simulations very well.

\end{document}